\documentclass[11pt]{article}
\usepackage{moriond,epsfig}

\def\Journal#1#2#3#4{{#1} {\bf #2}, #3 (#4)}

\def\NPB{{\em Nucl. Phys.} B}
\def\PLB{{\em Phys. Lett.}  B}

\def\ZPC{{\em Z. Phys.} C}
\def\CPC{{\em Comp. Phys. Comm. }}

\def\be{\begin{equation}}
\def\ee{\end{equation}}
\def\bc{\begin{center}}
\def\ec{\end{center}}
\def\bea{\begin{eqnarray}}
\def\eea{\end{eqnarray}}

\newcommand{\Afbol}{\mathrm{A_{FB}^{0,l}}}
\newcommand{\Afbob}{\mathrm{A_{FB}^{0,b\bar{b}}}}
\newcommand{\Afbb}{\mathrm{A_{FB}^{b\bar{b}}}}
\newcommand{\Afboc}{\mathrm{A_{FB}^{0,c\bar{c}}}}
\newcommand{\ALR}{\mathrm{A^0_{LR}}}
\newcommand{\ALRFB}{\mathrm{A^{LR}_{FB}}}
\newcommand{\sineff}{\mathrm{sin^2\theta^{lept}_{eff}}}
\newcommand{\sinw}{\mathrm{sin^2\theta_W}}

\newcommand{\mt}{\mathrm{m_{top}}}
\newcommand{\mw}{\mathrm{M_{W}}}
\newcommand{\mh}{\mathrm{M_{H}}}
\newcommand{\mz}{\mathrm{M_{Z}}}
\newcommand{\Dalhad}{\mathrm{\Delta\alpha^5_{had}}}
\newcommand{\alqed}{\mathrm{\alpha_{QED}}}
\newcommand{\als}{\mathrm{\alpha_s}}
\begin{document}
\vspace*{4cm}
\title{Electroweak Results from LEP and SLC and Tests of the Standard Model}

\author{E.TOURNEFIER }

\address{ISN Grenoble, 53, avenue des Martyrs, \\
  38026 Grenoble, France}

\maketitle\abstracts{
An update of the electroweak measurements at LEP and SLC is presented.
These measurements are used to perform precise tests of the Standard Model.
A constraint on the Standard Model Higgs mass is obtained when the direct measurements of 
$\mt$ and $\mw$ are included in the fit.
A combination with the direct Higgs search  is also shown.
}
\section{Introduction}\label{intro}
The precise electroweak measurements which have been performed at LEP and at SLC
allow to make precise tests of the Standard Model  and to constrain the Higgs mass.\\
At the Z resonance the cross sections  and the asymmetries of the process
$e^+e^-\rightarrow Z,\gamma \rightarrow f\bar{f}$
are sensitive to $\rm{m_t^{2}}$, $\als$ and $\rm{Log(\mh)}$ through radiative corrections.
The electroweak corrections to $e^+e^-\rightarrow f\bar{f}$ lead to the running of the electromagnetic coupling constant
$\alqed$ and corrections to the coupling constants of the Z to fermions.
These corrections are absorbed in the definition of the effective electroweak mixing angle
$\sineff$ and of $\bar{\rho}$:
$\mathrm{\sineff=(1+\Delta \kappa)\sin^2\theta_W}$ , 
$\mathrm{\bar{\rho}=1+\Delta \rho}$ 
where $\mathrm{\sin^2\theta_W=1-{M_W^2}/{M_Z^2}}$ and 
$\mathrm{\rho={M_W^2}/(M_Z^2 cos^2\theta_W)}=1$.
\\
The measurement of the asymmetries determines the values of 
${A}_l={\frac{2 g_{V}/g_{A}}{1+(g_{V}/g_A)^2}}$
which are converted into the  effective electroweak mixing angle 
$\sineff = \frac{1}{4}(1-\frac{g_V}{g_A})$, one of the most sensitive 
variables to the Higgs mass. On the other hand 
the measurement of the cross sections allows the determination of $\bar{\rho}$ which 
is more sensitive to the top mass.\\
The W mass also includes radiative corrections:
$\mathrm{M_W^2} = \frac{\pi \alpha}{\sqrt{2}{\sinw}{G_f}}(1+{\Delta r})$ with
${\Delta r = \Delta \alpha + \Delta r_{W}}$.
%
The values of these corrections, $\Delta \rho$, $\Delta \kappa$ and 
$\Delta r_{W}$ depend quadratically on $\mt$ and only 
logarithmically on $\mh$, leading to a much weaker constraint on $\mh$ than on $\mt$.\\
\section{The electroweak measurements}\label{EWmeas}
In this section the status of the main electroweak measurements
used in the fit to the Standard Model is given as well as the most significant new inputs.
\subsection{Status of the  measurements}\label{status}
The main electroweak measurements used in the fit are
\begin{itemize}
\item LEP1 and SLC electroweak measurements at the Z resonance:\\
The Z lineshape parameters from LEP1, the Z mass $\rm{M_Z}$, the 
Z width $\rm{\Gamma_Z}$, the hadronic pole cross section 
$\rm{\sigma^0_{had}}$, $\rm{R_l}=\Gamma_{\rm{hadrons}}/\Gamma_l$ and 
the forward-backward leptonic asymmetries
$\Afbol=\frac{3}{4}{\cal A}_e{\cal A}_l$  are final~\cite{Zcomb}.\\
The $\tau$ polarisation, ${\cal P}_{\tau}$ at LEP1 is also final.
${\cal A}_e$ and ${\cal A}_{\tau}$ are derived from this measurement. \\
The measurement of the left-right asymmetry $\ALR= {\cal A}_e$ and of 
the leptonic left-right forward-backward asymmetries $\ALRFB$ at SLC are final.\\
LEP1 and SLC also provide measurements of the Z decay fractions into $b$ 
and $c$ quarks  $\mathrm{R_b^0}$, $\mathrm{R_c^0}$. The $b\bar{b}$ and 
$c\bar{c}$ asymmetries 
$\Afbob$, $\Afboc$ as well as the quark charge asymmetry $<\mathrm{Q_{FB}}>$ 
are determined at LEP1 while $A_b$ and $A_c$ are measured at SLC.
ALEPH and DELPHI have significantly improved
their $\Afbob$ measurement~\cite{ALEPHAfbb,DELPHIAfbb} 
(see  section~\ref{afbb}).
SLD has updated its heavy flavour results~\cite{deGroot}. 
\item LEP2 and $\rm{p\bar{p}}$ colliders  measurements of the W mass: $\mw$
 from LEP
includes the data taken in 2000 for ALEPH and L3~\cite{Wmass}, this will be discussed in
section~\ref{mw}.
\item The top mass measurement from CDF and D0 which is final.
\item The determination of $\sinw$ by NuTeV.
\item Another important input used in the fit is the QED coupling constant at the Z mass
$\alqed(\rm{M_Z^2})$. 
New low energy $e^+e^-$ data taken by BES~\cite{BES} at BEPC
have been used to obtain a new experimental determination of 
$\alqed(\rm{M_Z^2})$~\cite{alqedBP}
(see section~\ref{alqed}).
\end{itemize}
Details and references to these measurements can be found in Reference~\cite{LEPEW2000,LEPEWpage}.
\subsection{The most significant new inputs}\label{new}
\begin{figure}
  \begin{minipage}{.48\linewidth}
    \bc
    \mbox{\epsfig{file=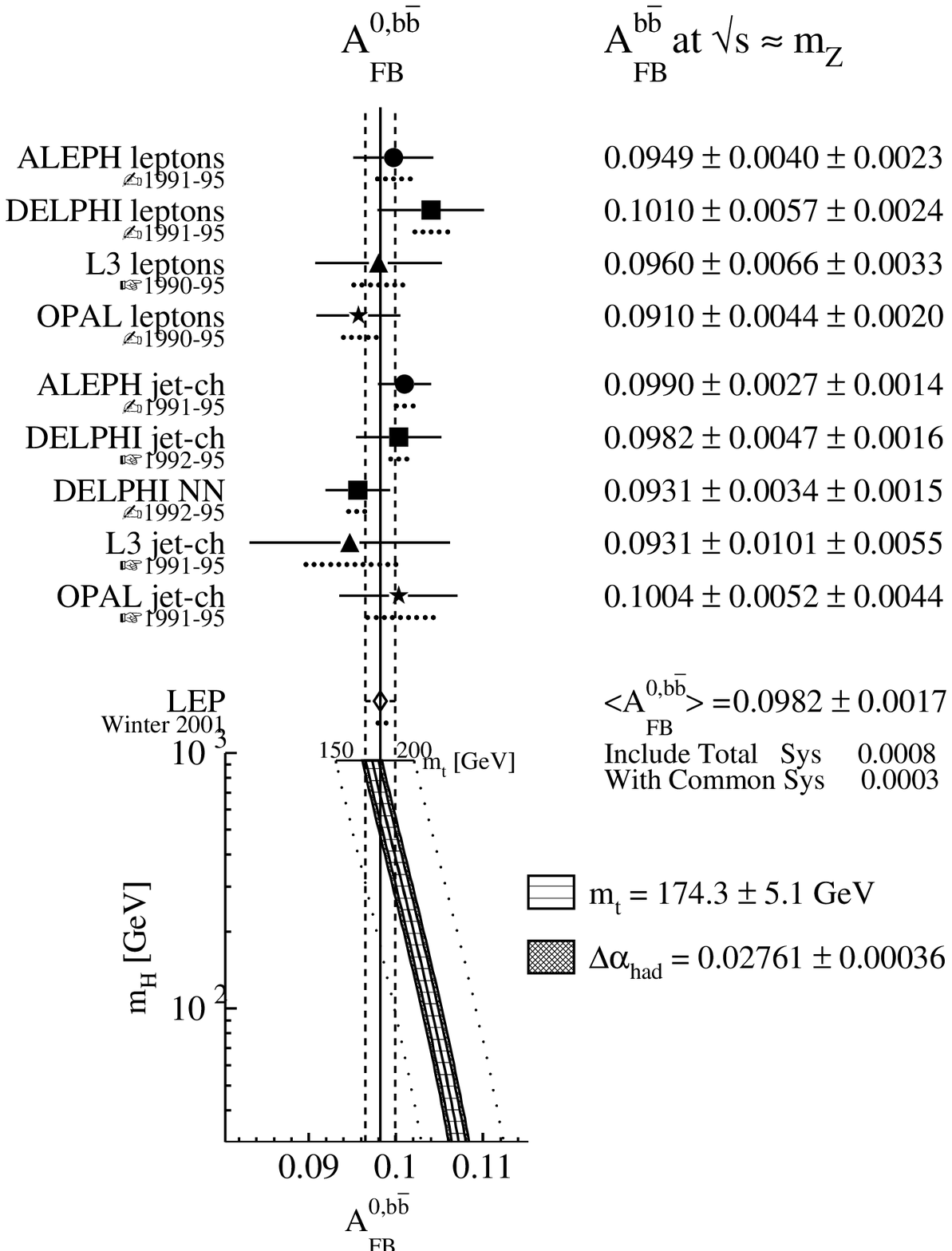,height=9.cm}}
    \ec
    \caption{Measurements of $\Afbob$ at LEP. The lower plot shows the prediction 
of the Standard Model as a function of $\mh$. The width of the band is due to the 
uncertainties on $\Dalhad$ and  $\mt$ added linearly.\label{afbbfig}}
  \end{minipage} \hspace{.5cm}
  \begin{minipage}{.48\linewidth}
    \bc
    \mbox{\epsfig{file=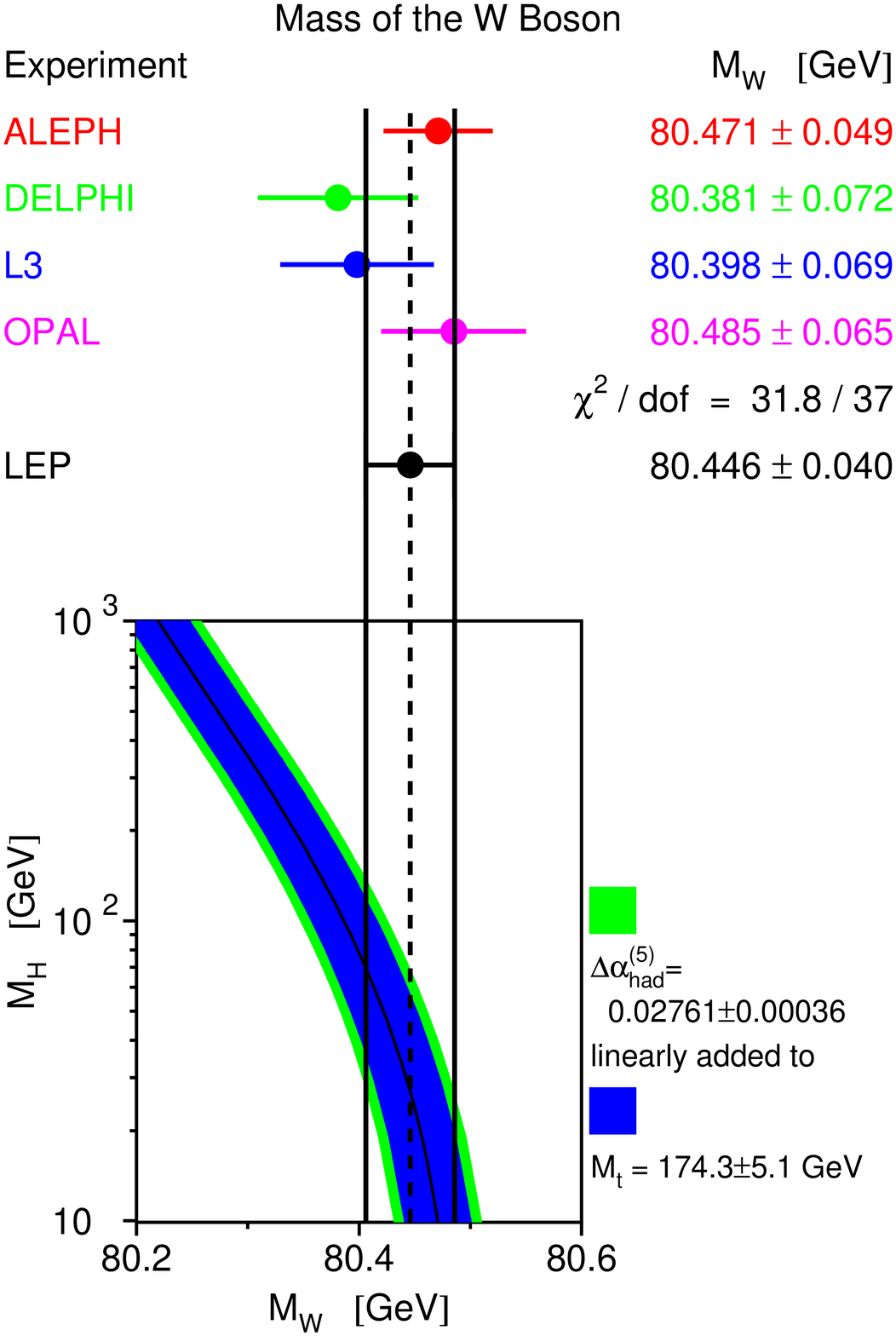,height=8.5cm}}
    \ec
    \caption{The measurement of $\mw$ at LEP. The lower plot shows the prediction 
of the Standard Model as a function of $\mh$. The width of the band is due to the 
uncertainties on $\Dalhad$ and  $\mt$  added linearly.\label{mwfig}}
  \end{minipage}
  \end{figure}

\underline{$\Afbob$}\label{afbb} \\\\
A new analysis~\cite{DELPHIAfbb} has been used by DELPHI leading to an 
improved determination of $\Afbb$. 
This analysis is based on a neural network to tag the b-charge
using the full available charge information from  vertex charge, 
jet charge and from identified leptons and hadrons. 
A double tag method is used to calibrate this neural network tag
on the data leading to a reduced systematic uncertainty. Note that this 
measurement is correlated with the measurement obtained with the {\it jet-charge} 
and with the {\it leptons} measurements shown in Figure~\ref{afbbfig}. 
The new value (refered to as DELPHI NN on Figure~\ref{afbbfig}) is 
\be
\Afbb(\sqrt{s}\simeq\mz) = 0.0931 \pm 0.0034 \pm 0.0015
\ee
ALEPH has improved its $\Afbb$ jet-charge measurement~\cite{ALEPHAfbb}.
A neural network has been used to tag b-events leading to a $30\%$ increase in 
 statistics while keeping the same purity. 
The jet-charge estimator has been improved, reducing the  mistag rate by $10\%$.
The systematic uncertainties are better controled by the use of 
double-tag methods for both flavour and charge tags.
With these improvements the systematic uncertainty has been reduced by a factor of 2 
with respect to the old value~\cite{ALEPHAfbbold}:
\bea
\Afbb(\sqrt{s}\simeq\mz) &= 0.0990 \pm 0.0027 \pm 0.0014 \\
\rm{old\hspace{1mm} value:\hspace{1mm}} 
\Afbb(\sqrt{s}\simeq\mz) &=0.1017 \pm 0.0038\pm 0.0032
\eea
Figure~\ref{afbbfig} shows all the $\Afbb$ measurements.\\\\
\underline{The W mass}\label{mw} \\\\
At LEP2 the W mass is determined~\cite{Wmass}  from the data  recorded at 
centre-of-mass energies $\sqrt{s}=161-209$~GeV.
Only ALEPH and L3 have analysed the year 2000 data and have an integrated 
luminosity of 700~pb$^{-1}$ per experiment. The DELPHI and OPAL results are 
based on only 450~pb$^{-1}$ per experiment.\\
Moreover ALEPH has done futher systematic studies~\cite{ALEPHWmass} 
leading to a significantly
reduced  uncertainty. The fragmentation uncertainty which is based on Monte 
Carlo comparisons is reduced from 30~MeV to 15~MeV.
The uncertainty arising from final state interaction between the two W's 
(Bose Einstein Correlation and Color Reconnection) has also been re-estimated.
\\
The LEP combined W mass obtained from the 
$\mathrm{q\bar{q}l\bar{\nu_l}}$ and $\mathrm{q\bar{q}q\bar{q}}$ channels
are consistent:
\begin{center}
$\mathrm{\Delta\mw(q\bar{q}q\bar{q} - q\bar{q}l\bar{\nu_l})} = +18\pm 46 \rm{MeV}$
\end{center}
showing no evidence for a bias arising from FSI effects.
The LEP W mass measurement is shown in Figure~\ref{mwfig}:
\be
\mw = 80.446 \pm 0.026(\rm{stat}) \pm 0.030(\rm{syst}) \rm{ GeV}
\ee 
The combination with the CDF, D0 and UA2 measurements gives:
\be
\mw = 80.448 \pm 0.034 \rm{ GeV}
\ee 
\underline{ $\alqed(\rm{M_Z^2})$}\label{alqed}\\\\
As pointed out in section~\ref{intro} the value of the QED coupling constant 
at $\sqrt{s}=\mz$, $\alqed(\rm{M_Z^2})$ is needed in the fits. 
The running of $\alqed$ is given by:
\be
\alpha(s) = \frac{\alpha(0)}{1 - {\Delta\alpha_l(s)} - {\Dalhad(s)} - {\Delta\alpha_{top}(s)}} 
\ee
$\Delta\alpha_l(s)$ and $\Delta\alpha_{top}(s)$ are well known while $\Dalhad(s)$ involves hadron loops at low energy and therefore non perturbative QCD.
This can nevertheless be experimentally determined using the low energy 
 $e^+e^-$ data since $\Dalhad(s)$ is related to 
${R_{had} = \frac{\sigma({e^+e^-\rightarrow\mathrm{ hadrons}})}
{\sigma({e^+e^-\rightarrow \mu^+\mu^-})}}$ via a dispersion integral.\\
In the previous determinations~\cite{alqedEJ} of $\Dalhad(s)$
 the dominant error came from data taken in the range
$2<\sqrt{s}<5$~GeV. The error in this energy range has been reduced
by a factor of more than 2 using new  $e^+e^-$ data from the BES experiment~\cite{BES}.\\
A new determination of $\Dalhad$ has been done~\cite{alqedBP} using 
only experimental data below 12~GeV and third order QCD above, leading to
\begin{equation}
  \Dalhad(\mz) = 0.02761 \pm 0.00036
\end{equation}
The error has been reduced by almost  a factor 2 with respect to the previous
value used in the electroweak fit~\cite{alqedEJ}:
$\Dalhad(\mz) = 0.02804 \pm 0.00065$.
Another determination including the BES data but more theoretical inputs in the low 
energy region~\cite{alqedMOR}, $\Dalhad(\mz) = 0.02738 \pm 0.00020$, will also be used 
in the fit for comparison.
%
\subsection{Sensitivity  to the Higgs mass}\label{sensitivity}
\begin{figure}
  \begin{minipage}{.47\linewidth}
    \bc
    \mbox{\epsfig{file=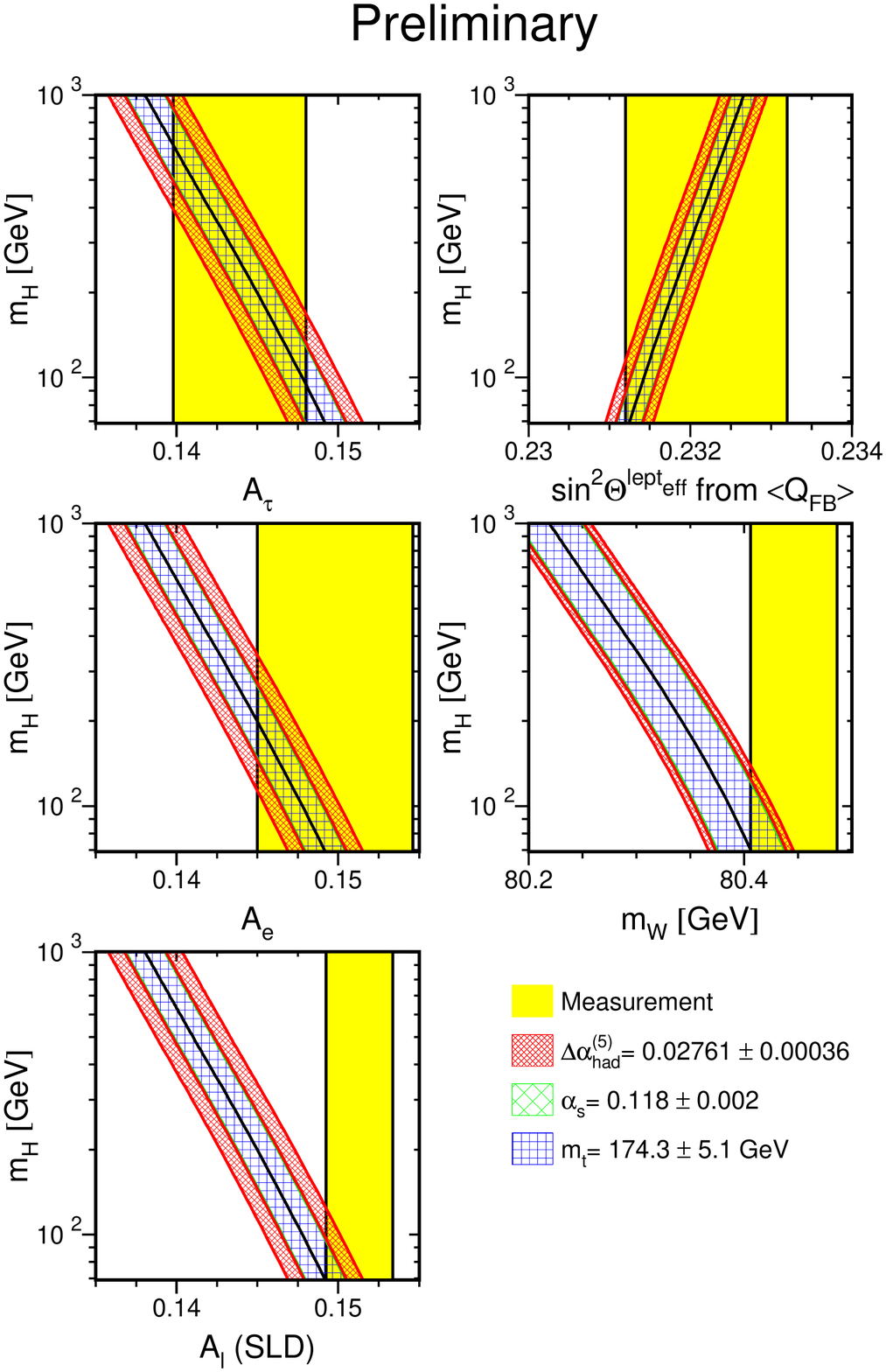,height=9.7cm}}  
    \ec
    \caption{Sensitivity of some asymmetries and of $\mw$ to the 
      Higgs mass. The width of the Standard Model band gives the 
      uncertainty arising from the precision on $\Dalhad$ and $\mt$ and $\alpha_s$
      added linearly.
      \label{sensitivity1}}
  \end{minipage}\hspace{.5cm}
  \begin{minipage}{.47\linewidth}
    \bc
    \mbox{\epsfig{file=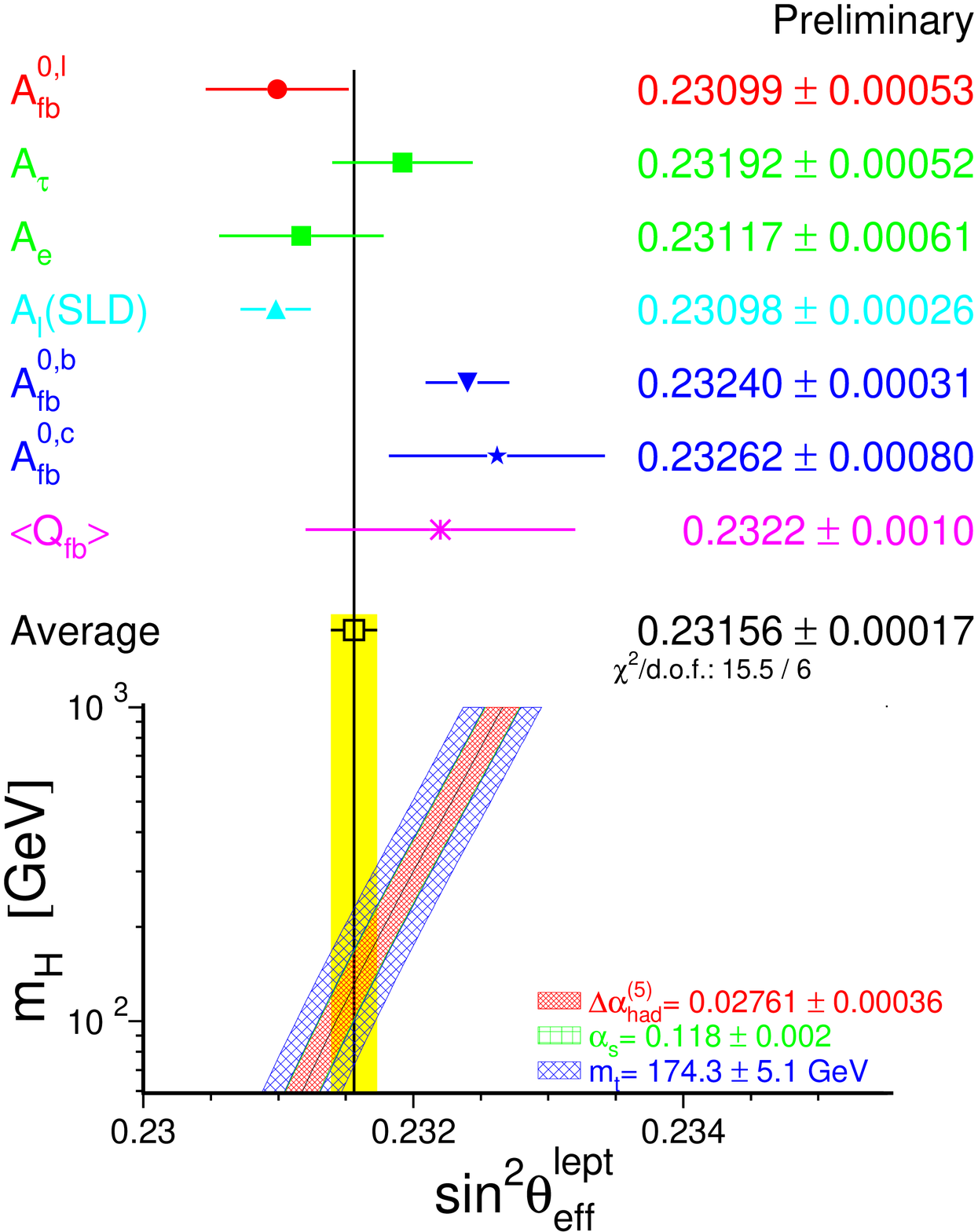,height=9.cm}}  
    \ec
    \caption{Detemination of $\sineff$ from the asymmetry measurements.
      The Standard Model prediction as a function of $\mh$ is also shown.
      The width of this prediction gives the 
      uncertainty arising from the band on $\Dalhad$ and $\mt$ and $\alpha_s$
      added linearly.
      \label{sineff}}
  \end{minipage}
\end{figure}

Figure~\ref{sensitivity1} shows the sensitivity of some asymmetries
and of the W mass to the Higgs mass. The experimental measurements are shown as well
as the Standard Model prediction. The width of the Standard Model band 
shows the uncertainty arising from the precision on $\Dalhad$, $\mt$ and $\als$.
The asymmetries are very sensitive to the Higgs mass. The W mass
is also  sensitive but it is very dependent on $\mt$.
\\
All the asymmetry measurements can be converted into the measurement of the single 
parameter $\sineff$ (Figure~\ref{sineff}).
The combination of these measurements gives
\be
\sineff = 0.23156 \pm 0.00017
\ee
The $\chi^2$ of the fit is bad: $\chi^2/d.o.f. = 15.5/6$. 
This reflects the fact that the combined value of $\sineff$ obtained from
the leptonic asymmetries is  3.6$\sigma$ apart from that obtained
with the quark asymmetries. This effect is mainly caused by the 2 most precise
measurements, $\Afbob$ at LEP and ${A_l}$ from SLD.
Since the previous combination, $\Afbob$ has been 
more precisely measured as explained in section~\ref{afbb} and its value has slightly
decreased, and so prefering a high Higgs mass (around 600~GeV).
On the contrary the leptonic asymmetries prefer a  light Higgs 
(around 60~GeV). 
This dispersion is interpreted here as a fluctuation in one or more of the measurements.
\section{Test of the Standard Model}\label{SMtest}
\begin{figure}
  \begin{minipage}{.47\linewidth}
    \bc
    \mbox{\epsfig{file=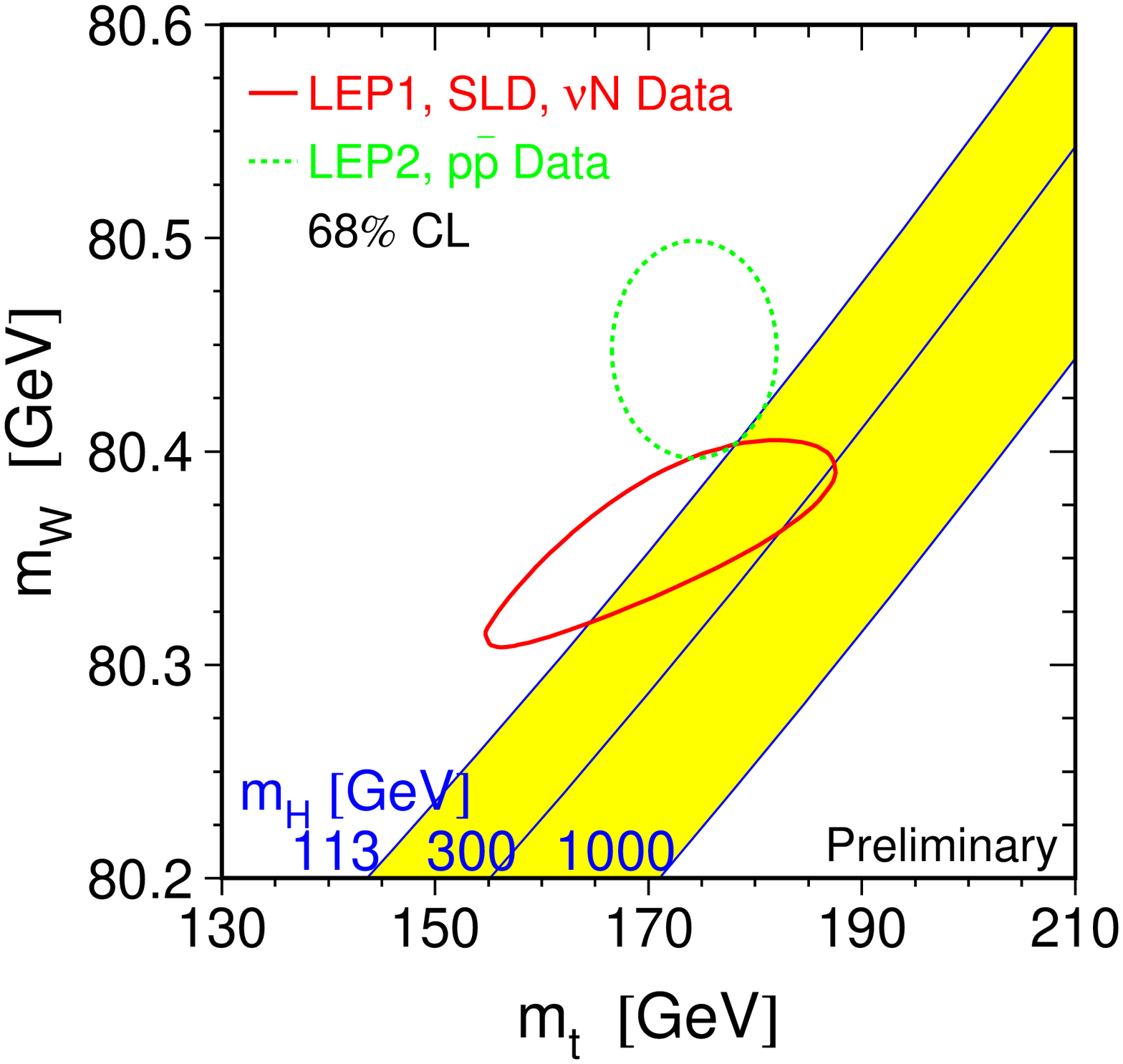,height=8cm}}  
    \ec
    \caption{Indirect determination of $\mt$ and $\mw$ (full line)
      compared to the direct measurements (dotted line). 
      The 68$\%$ C.L. contours are shown.
      The band shows the Standard Model prediction for a Higgs mass 
      ranging from 113~GeV to 1~TeV.
      \label{mt_mw}}
  \end{minipage}\hspace{.5cm}
  \begin{minipage}{.47\linewidth}
    \bc
    \mbox{\epsfig{file=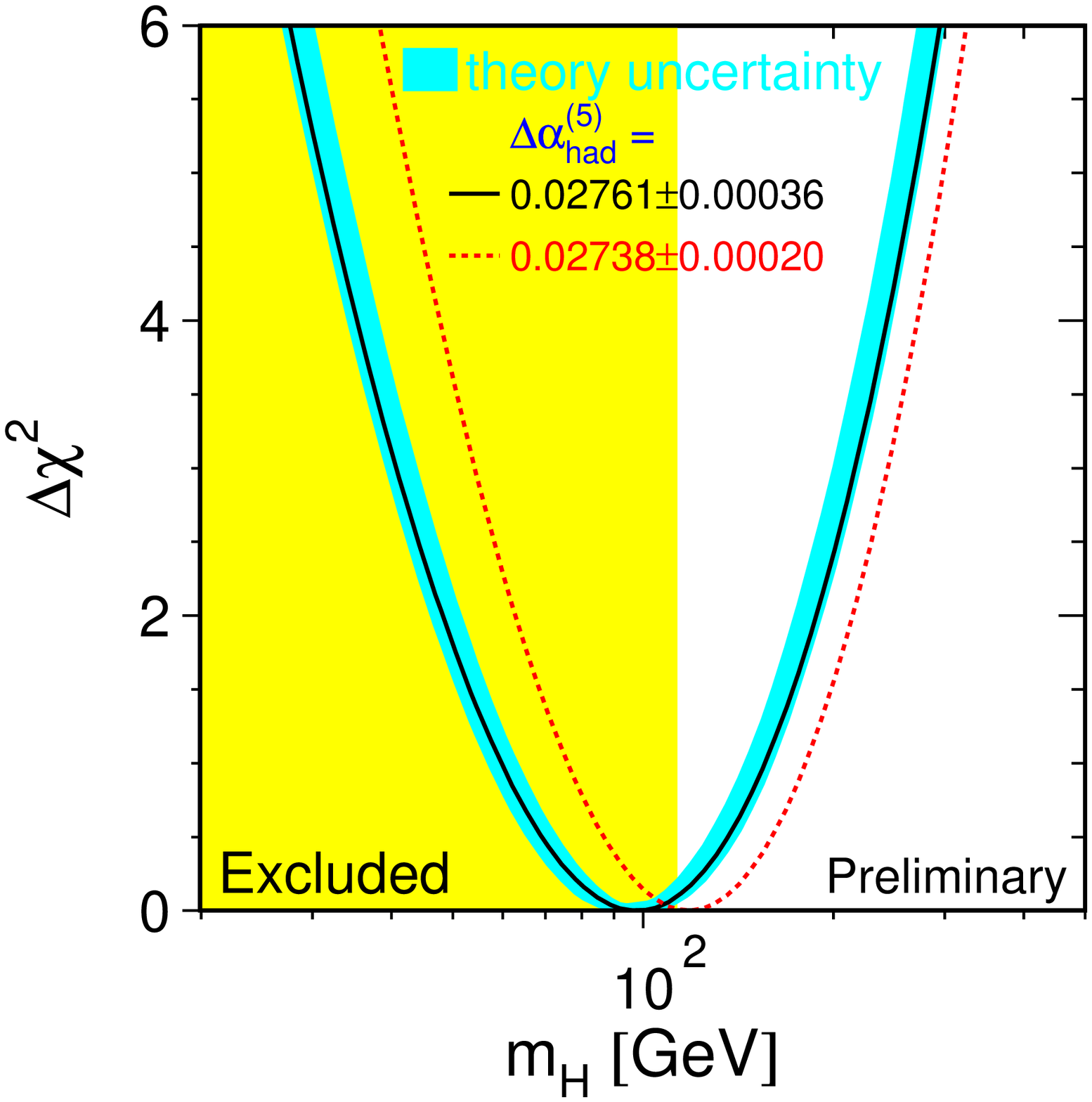,height=8cm}}  
    \ec
    \caption{$\Delta \chi^2 = \chi^2 - \chi^2_{min}$ as a function of $\mh$ 
      for  the global fit to the Standard Model. 
      The region excluded by the direct searches at LEP 
      is also shown.
      \label{globalfit}}
  \end{minipage}
\end{figure}

In the following the ZFITTER~\cite{ZFITTER} and TOPAZ0~\cite{TOPAZ0} 
programs are used for all the fits.
Using all the measurements discussed in section~\ref{status}
except the direct measurement of $\mw$ and $\mt$ a fit to the Standard Model 
is performed  to 
obtain an indirect determination of $\mw$ and $\mt$ and a constraint on $\mh$.
Figure~\ref{mt_mw} shows the result of that fit as well as the direct 
measurements (the 68$\%$ C.L.  contours are shown). This indirect determination
gives
\bea
\mt =& 168.3^{+11.9}_{-9.3} \rm{ GeV}\\
\mw =& 80.357\pm 0.033 \rm{ GeV}\\
\mathrm{Log}(\mh) =& 1.94^{+0.37}_{-0.30}\\
\mh =& 87^{+119}_{-43} \rm{ GeV}
\eea
in agreement with the direct measurements:
\bea
\mt =& 174.3 \pm 5.1 \rm{ GeV}\\
\mw =& 80.448\pm 0.034 \rm{ GeV}
\eea
The Standard Model prediction is also shown in Figure~\ref{mt_mw} showing that a low
Higgs mass is prefered by both the direct and indirect $\mw$ and $\mt$ values.
%
\section{Constraint on the Higgs mass}\label{Higgsmass}
\subsection{The global fit}\label{fit}
The agreement between the direct and the indirect determination of 
$\mw$ and $\mt$ (section~\ref{SMtest}) shows the consistency of the  Standard Model.
These direct measurements are then used  to obtain a better 
constraint on the Higgs mass.
Using the new value of $\Dalhad$ discussed in 
section~\ref{alqed} the result of the fit is:
\bea
\mathrm{Log}(\mh) =& 1.99\pm0.21\\
\mh =& 98^{+58}_{-38} \rm{ GeV}\\
\mt =& 175.7\pm4.4 \rm{ GeV}\\
\mw =& 80.393\pm 0.019 \rm{ GeV}
\eea
leading to an upper limit on the Higgs mass: $\mh< 212$~GeV at 95\% C.L.
Again the $\chi^2$ is bad, $\chi^2/d.o.f.=25/15$ which corresponds 
to a probability of only $4\%$. This is simply a reflection of the  the disagreement 
between the asymmetry measurements  already discussed in section~\ref{sensitivity}.\\
Figure~\ref{globalfit} shows the $\Delta\chi^2= \chi^2-\chi^2_{min}$ as a 
function of the Higgs mass. The dotted line shows the result of the fit using 
a more theory driven determination of $\Dalhad$~\cite{alqedMOR} which also 
includes the new BES data.\\
The new value of $\Dalhad$ results in a shift of the prefered Higgs mass of about 
$+35$~GeV and a significant reduction in the error: the error on  Log($\mh$)
arising from $\Dalhad$ has decreased from 0.2 to 0.1. This is no longer 
the single dominant error, but still one of the limiting errors.
\subsection{The uncertainty on $\mh$}\label{error}
In order to determine which measurements need to be improved  to  better 
constrain the Higgs mass the error on Log($\mh$) can be broken down into
 the different sources. For this purpose only the two most powerful
variables, $\sineff$ and $\mw$, are used separately.
The parametrisation of the $\sineff$ and $\mw$ dependance with 
$\rm{m_{top}^2}$, Log($\mh$) and $\Dalhad$ given in Ref.~\cite{Degrassi}
is used to propagate the experimental errors.
Using $\sineff$ alone the uncertainty on Log($\mh$) is 
\begin{equation}
  \rm{\delta {Log}({\mh}) = \pm0.14(\delta sin^2\theta_{lept}^{eff})
  \mp0.10(\delta \Dalhad)
  \pm0.13(\delta \mt) = {\pm 0.22}}
\end{equation}
and using $\mw$ alone:
\begin{equation}
  \rm{\delta {Log}({\mh}) = \mp0.24(\delta M_W) 
    \mp0.05(\delta \Dalhad) 
    \pm0.26(\delta \mt) = {\pm 0.36}}
\end{equation}
The error arising from $\Dalhad$ is no more the dominant one and with the 
present uncertainty on $\mw$ and on $\mt$  $\sineff$ is more powerful than
$\mw$. The large dependence of $\mw$ on $\mt$ limits the power of
$\mw$ in constraining $\mh$. Therefore one also needs to improve the precision of 
$\mt$ in order to increase the power of $\mw$.
Assuming that at the end of Run IIa of the  Tevatron~\cite{Glenzinski}
 the error on $\mt$ is reduced
 to 2.5~GeV and on $\mw$ to 25~MeV (LEP 2 and Tevatron combined) then
one would get using $\sineff$ alone:
\begin{equation}
  \rm{\delta {Log}({\mh}) = \pm0.14(\delta sin^2\theta_{lept}^{eff})
  \mp0.10(\delta \Dalhad)
  \pm0.07(\delta \mt) = {\pm 0.19}}
\end{equation}
and using $\mw$ alone:
\begin{equation}
  \rm{\delta {Log}({\mh}) = \mp0.14(\delta M_W) 
    \mp0.05(\delta \Dalhad) 
    \pm0.12(\delta \mt) = {\pm 0.19}}
\end{equation}
$\mw$ would then be as powerful as $\sineff$ for constraining $\mh$.
Note that these numbers are obtained assuming that the value of $\mh$ is of 
the order of 100~GeV.
\subsection{Combination with the direct search}\label{comb}
\begin{figure}
  \begin{center}
    \mbox{\epsfig{file=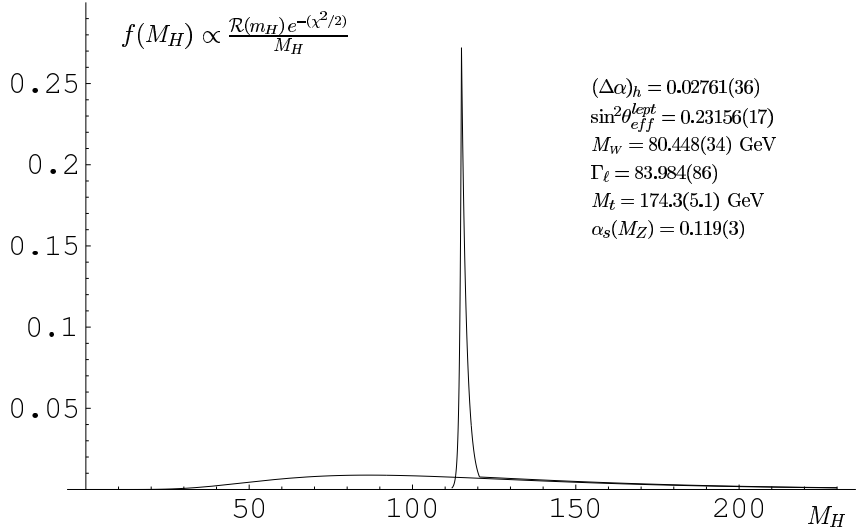,height=7.cm}}  
  \end{center}
  \caption{Probability density function $f(\mh)$. The lower curve shows the 
    indirect measurements alone and the curve with the spike at $\sim$115~GeV 
    shows the combination with the direct search.
    \label{combfig}}
\end{figure}

In Figure~\ref{globalfit} the lower limit on the Higgs mass obtained from the 
direct searches at LEP2 is also shown, but this information is not
used in the fit. 
The likelihood ${\cal R}(\mh$) obtained from the direct
 searches~\cite{LEPC} is combined with the 
$\chi^2$ probability obtained from the indirect measurements in Ref.~\cite{Degrassi_comb}.
${\cal R}(\mh$) includes the information from the excess of events observed in the year 
2000 at a mass around 115~GeV.  This combination uses a Baysian approach, 
assuming a uniform  prior in Log($\mh$). The probability density function  
$f(\mh) \propto \frac{{\cal R}(\mh ) e^{(-\chi^2/2)}}{\mh}$ is shown in 
Figure~\ref{combfig}.
The  spike at $\mh \simeq 115$~GeV is due to the excess of events in the direct search. 
The effect of this excess is to concentrate most of the probability around 115~GeV.
About 50$\%$ of the probability is contained between a 
mass of 113~GeV and 120~GeV.
Note also that the $95\%$ upper limit goes up by about 20~GeV when the direct 
search is taken into account.
\section{Conclusion}
LEP1 and SLD results are  final except the heavy flavour results.
The values of $\sineff$ extracted from the leptonic asymmetries and from the 
quark asymmetries are 3.6$\sigma$ apart. This effect is interpreted here as 
a fluctuation in one or more of the measurements.\\
The new BES data lead to a significant improvement in the determination of $\Dalhad$
which used to be the dominant source of error in the electroweak fit.\\
The full LEP2 data set is not yet analysed for the W mass measurement.
Moreover systematic uncertainties
should be reduced with studies of the full data set.\\
Thanks to the statistics which will be accumulated during  RunIIa at the Tevatron
the uncertainty on the top mass and on the W mass will be significantly reduced
allowing us to make more precise tests of the Standard Model and to  
constrain better the Higgs mass.
\section*{Acknowledgments}
I would like to thank the LEP Electroweak Working group for providing 
me with the results and the plots. I also thank G. Degrassi for kindly 
sending me his results.


\section*{References}
\begin{thebibliography}{99}
\bibitem{Zcomb}The LEP~Collaborations and  the Line Shape Sub-group of the LEP
 Electroweak Working Group 
{\em CERN-EP-2000-153}, to appear in  {\em Phys. Rep.}
\bibitem{ALEPHAfbb}The ALEPH Collaboration, Measurement of $\Afbb$ using 
Inclusive b-hadron Decays, {\em ALEPH-CONF/2001-020}.
\bibitem{DELPHIAfbb}The DELPHI Collaboration, Determination of $\Afbb$ using 
inclusive charge reconstruction and lifetime tagging at LEP~1,
{\em DELPHI/2001-027 CONF 468}.
\bibitem{deGroot}N. de Groot, in these proceedings.
\bibitem{Wmass}The LEP~Collaborations and the LEPEW Working Group,
{\em LEPEWWG/MASS/2000-01}.
\bibitem{BES} The BES Collaboration, J.Z. Bai {\it et al}, hep-ex/01023003 and
G. Huang, contribution to these proceedings.
\bibitem{alqedBP}H. Burkhardt and B. Pietrzyk, LAPP-EXP 2001-03, 
  accepted by {\PLB}.
\bibitem{LEPEW2000}The LEP Collaborations, the LEP Electroweak Working Group
and the SLD Heavy Flavour and Electroweak Groups,{\em CERN-EP/2001-021}.
\bibitem{LEPEWpage} The LEP Electroweak Working Group: 
http://lepewwg.web.cern.ch/LEPEWWG/
\bibitem{ALEPHAfbbold}The ALEPH Collaboration, Determination of $\Afbb$ using
Jet Charge Measurements in Z Decays, \Journal{\PLB}{426}{217}{1998}.
\bibitem{ALEPHWmass}The ALEPH Collaboration, Measurement of the W Mass and 
Width in $e^+e^-$ Collisions  at $\sqrt{s}$ between 192 and 208~GeV, 
{\em ALEPH-CONF/2001-017}.
\bibitem{alqedEJ}S. Eidelmann and F. Jegerlehner, \Journal{\ZPC}{67}{585}{1995}.
\bibitem{alqedMOR}A. Martin, J. Outwhaite and M.G. Ryskin, \Journal{\PLB}{492}{69}{2000}.
\bibitem{ZFITTER}D. Bardin {\it et al.}, \Journal{\ZPC}{44}{493}{1989};
\Journal{C.P.C}{59}{303}{1990}; ZFITTER v.6.21:DESY 99-070(1999), hep-ph/9908433 to appear in \CPC.
\bibitem{TOPAZ0}G. Montagna {\it et al.}, \Journal{\CPC}{117}{278}{1999}.
\bibitem{Degrassi}G. Degrassi, P. Gambino, \Journal{\NPB}{567}{3}{2000}.
\bibitem{Glenzinski}D. Glensinski, contribution to  these proceedings.
\bibitem{LEPC}P. Igo-Kemenes, Talk given at the LEPC of November 3, 2000.
\bibitem{Degrassi_comb}G. Degrassi,Talk presented at 50 Years of Electroweak 
Physics,New York, 27-28 Oct 2000 hep-ph/0102137.
\end{thebibliography}
\end{document}